\documentclass[a4paper,11pt]{article}
\usepackage{pos}
\title{Lattice QCD study of antiheavy-antiheavy-light-light tetraquarks based on correlation functions with scattering interpolating operators both at the source and at the sink}
\ShortTitle{Lattice QCD study of antiheavy-antiheavy-light-light tetraquarks}

\author*[a,b]{Marc Wagner}
\author[c,d]{Constantia Alexandrou}
\author[d]{Jacob Finkenrath}
\author[e]{Theodoros Leontiou}
\author[f]{Stefan Meinel}
\author[a]{Martin Pflaumer}

\affiliation[a]{Johann Wolfgang Goethe-Universit\"at Frankfurt am Main, Institut f\"ur Theoretische Physik, Max-von-Laue-Stra{\ss}e 1, D-60438 Frankfurt am Main, Germany}

\affiliation[b]{Helmholtz Research Academy Hesse for FAIR, Campus Riedberg, Max-von-Laue-Stra{\ss}e 12, \\ D-60438 Frankfurt am Main, Germany}

\affiliation[c]{Department of Physics, University of Cyprus, P.O.\ Box 20537, 1678 Nicosia, Cyprus}

\affiliation[d]{Computation-based Science and Technology Research Center, The Cyprus Institute, 20 Kavafi Street, 2121 Nicosia, Cyprus}

\affiliation[e]{Department of Mechanical Engineering, Frederick University, 1036 Nicosia, Cyprus}

\affiliation[f]{Department of Physics, University of Arizona, Tucson, AZ 85721, USA}

\emailAdd{mwagner@itp.uni-frankfurt.de}
\emailAdd{c.alexandrou@cyi.ac.cy}
\emailAdd{j.finkenrath@cyi.ac.cy}
\emailAdd{t.leontiou@frederick.ac.cy}
\emailAdd{smeinel@arizona.edu}
\emailAdd{pflaumer@itp.uni-frankfurt.de}

\abstract{We present first results of a recently started lattice QCD investigation of antiheavy-antiheavy-light-light tetraquark systems including scattering interpolating operators in correlation functions both at the source and at the sink. In particular, we discuss the importance of such scattering interpolating operators for a precise computation of the low-lying energy levels. We focus on the $\bar b \bar b u d$ four-quark system with quantum numbers $I(J^P) = 0(1^+)$, which has a ground state below the lowest meson-meson threshold. We carry out a scattering analysis using L\"uscher's method to extrapolate the binding energy of the corresponding QCD-stable tetraquark to infinite spatial volume. Our calculation uses clover $u$, $d$ valence quarks and NRQCD $b$ valence quarks on gauge-link ensembles with HISQ sea quarks that were generated by the MILC collaboration.}

\FullConference{%
The 39th International Symposium on Lattice Field Theory,\\
8th-13th August, 2022,\\
Rheinische Friedrich-Wilhelms-Universit\"at Bonn, Bonn, Germany
}

\newcommand{\C}{\mathcal{C}}
\newcommand{\xb}{\mathbf{x}}
\newcommand{\yb}{\mathbf{y}}
\newcommand{\op}{\mathcal{O}}

\begin{document}

\maketitle

% ********************

\section{Introduction}

We report on a recently started lattice QCD project in which we aim to study possibly existing heavy-heavy-light-light tetraquark resonances. In the following, we focus on the $\bar{b}\bar{b} u d$ tetraquark with quantum numbers $I(J^P) = 0(1^+)$, which is theoretically simpler compared to other tetraquark candidates because it is QCD-stable. This tetraquark is the counterpart of the $\bar c \bar c u d$ tetraquark $T_{cc}$ recently discovered by LHCb \cite{LHCb:2021vvq,LHCb:2021auc}.

In the past couple of years, several independent lattice QCD studies of $\bar{b} \bar{b} q q$ and $\bar{b} \bar{c} q q$ systems ($q$ denotes a light $u$, $d$ or $s$ quark) were published. These computations employed either exclusively local four-quark interpolating operators \cite{Francis:2016hui,Francis:2018jyb,Junnarkar:2018twb,Hudspith:2020tdf,Mohanta:2020eed} or local and scattering four-quark interpolating operators, but the latter only at the sink \cite{Leskovec:2019ioa,Meinel:2022lzo}.

In the work presented here, we include scattering interpolating operators both at the source and at the sink. This allows a more precise determination of finite-volume energy levels not only for bound states, but also for scattering states. This is particularly important for $\bar Q \bar Q q q$ systems, where bound states and scattering states are very close, or where bound states do not exist but resonances might exist. An example is a future full lattice QCD investigation of a possibly existing $\bar b \bar b u d$ tetraquark resonance with $I(J^P) = 0(1^-)$ proposed in Ref.\ \cite{JH2022}.

% ********************

\section{Interpolating operators}

To study the $\bar b \bar b u d$ four-quark system with quantum numbers $I(J^P) = 0(1^+)$, we use local interpolating operators
\begin{eqnarray}
 & & \hspace{-0.7cm} \op_1 = \op_{[B B^\ast](0)} = \sum_{\xb}	\bar{b} \gamma_5 d(\xb) \, \bar{b}\gamma_j u(\xb) - (d \leftrightarrow u),	\\
 & & \hspace{-0.7cm} \op_2 = \op_{[B^\ast B^\ast](0)} = \epsilon_{j k l} \sum_{\xb}	\bar{b} \gamma_k d(\xb) \, \bar{b} \gamma_l u(\xb)	- (d \leftrightarrow u), \\
 & & \hspace{-0.7cm} \op_3 = \op_{[D d](0)} = \sum_{\xb}\bar{b}^a \gamma_j \C \bar{b}^{b,T}(\xb)\, d^{a,T} \C \gamma_5  u^b(\xb) - (d \leftrightarrow u),
\end{eqnarray}
and scattering interpolating operators
\begin{eqnarray}
 & & \hspace{-0.7cm} \op_4 = \op_{B(0) B^\ast(0)} = \bigg(\sum_{\xb} \bar{b} \gamma_5 d(\xb)\bigg) \, \bigg(\sum_{\yb}\bar{b} \gamma_j u(\yb)\bigg) - (d \leftrightarrow u), \\
 & & \hspace{-0.7cm} \op_5 = \op_{B^\ast(0) B^\ast(0)} = \epsilon_{j k l} \bigg(\sum_{\xb} \bar{b} \gamma_k d(\xb)\bigg) \, \bigg(\sum_{\yb} \bar{b} \gamma_l u(\yb)\bigg) - (d \leftrightarrow u) .
\end{eqnarray}
Here, $\C$ denotes the charge conjugation matrix, and upper indices $a$ and $b$ are color indices. For more details we refer to our previous work \cite{Leskovec:2019ioa}.

% ********************

\section{Lattice setup}

We use $2+1+1$-flavor HISQ gauge-link ensembles generated by the MILC collaboration \cite{MILC:2012znn} as summarized in Table~\ref{TAB001}.

\begin{table}[htb]
\begin{center}
\begin{tabular}{lccccc}\hline \hline
		Ensemble & $ a $ [fm] & $ N_s^3\times N_t $ & $ m_{\pi}^{\tiny (\textrm{sea})} $ [MeV] & $ m_{\pi}^{\tiny (\textrm{val})} $ [MeV] & $ N_\textrm{\tiny conf} $ \\\hline
		a12m310  & $0.1207(11)$ & $ 24^3 \times 64 $ & $ 305.3(4) $ & $ 310.2(2.8) $ & $ 1053 $\\
		a12m220S & $0.1202(12)$ & $ 24^3 \times 64 $ & $ 218.1(4) $ & $ 225.0(2.3) $ & $ 1020 $\\
		a12m220  & $0.1184(10)$ & $ 32^3 \times 64 $ & $ 216.9(2) $ & $ 227.9(1.9) $ & $ 1000 $\\
		a12m220L & $0.1189(09)$ & $ 40^3 \times 64 $ & $ 217.0(2) $ & $ 227.6(1.7) $ & $ 1030 $\\
		a09m310  & $0.0888(08)$ & $ 32^3 \times 96 $ & $ 312.7(6) $ & $ 313.0(2.8) $ & $ 1166 $\\
		a09m220  & $0.0872(07)$ & $ 48^3 \times 96 $ & $ 220.3(2) $ & $ 225.9(1.8) $ & $ \phantom{0}657 $\\ \hline \hline
\end{tabular}
\end{center}
\caption{\label{TAB001}Gauge-link ensembles ($a$: lattice spacing; $N_s$, $N_t$: number of lattice sites in spatial and temporal direction; $m_{\pi}^{\tiny (\textrm{sea})}$, $m_{\pi}^{\tiny (\textrm{val})}$: pion mass corresponding to light sea and light valence quarks; $N_\textrm{\tiny conf}$: number of gauge-link configurations used for computations).}
\end{table}

We use a mixed-action setup with Wilson-clover $u$ and $d$ valence quarks \cite{Bhattacharya:2015wna,Gupta:2018qil}. For the $b$ valence quarks we use lattice NRQCD \cite{HPQCD:2011qwj}.

Correlation functions are computed with point-to-all propagators if there is a local operator at the source. If there is a scattering operator at the source, we use stochastic timeslice-to-all propagators combined with the one-end trick (see e.g.\ Ref.\ \cite{Abdel-Rehim:2017dok}). Moreover, we use APE smearing for the gauge links and Gaussian smearing for the quark fields.

For the analysis of the correlation matrices we employ two independent methods: solving standard generalized eigenvalue problems (GEVP) as well as the Athens Model Independent Analysis Scheme (AMIAS) \cite{Alexandrou:2014mka}.

% ********************

\section{\label{SEC001}Effective masses from ``exclusively local'' versus ``local and scattering'' interpolating operators}

Based on previous lattice QCD computations \cite{Francis:2016hui,Junnarkar:2018twb,Mohanta:2020eed,Leskovec:2019ioa} we expect the ground state around $100 \, \textrm{MeV}$ below the $B B^\ast$ threshold (the lowest meson-meson threshold in this channel) representing the QCD-stable tetraquark. The first and second excitations in the finite spatial volume should be meson-meson scattering states resembling $B B^\ast$ and $B^\ast B^\ast$ close to the respective thresholds.

\begin{figure}[htb]
\begin{center}
\includegraphics[width=0.48\textwidth]{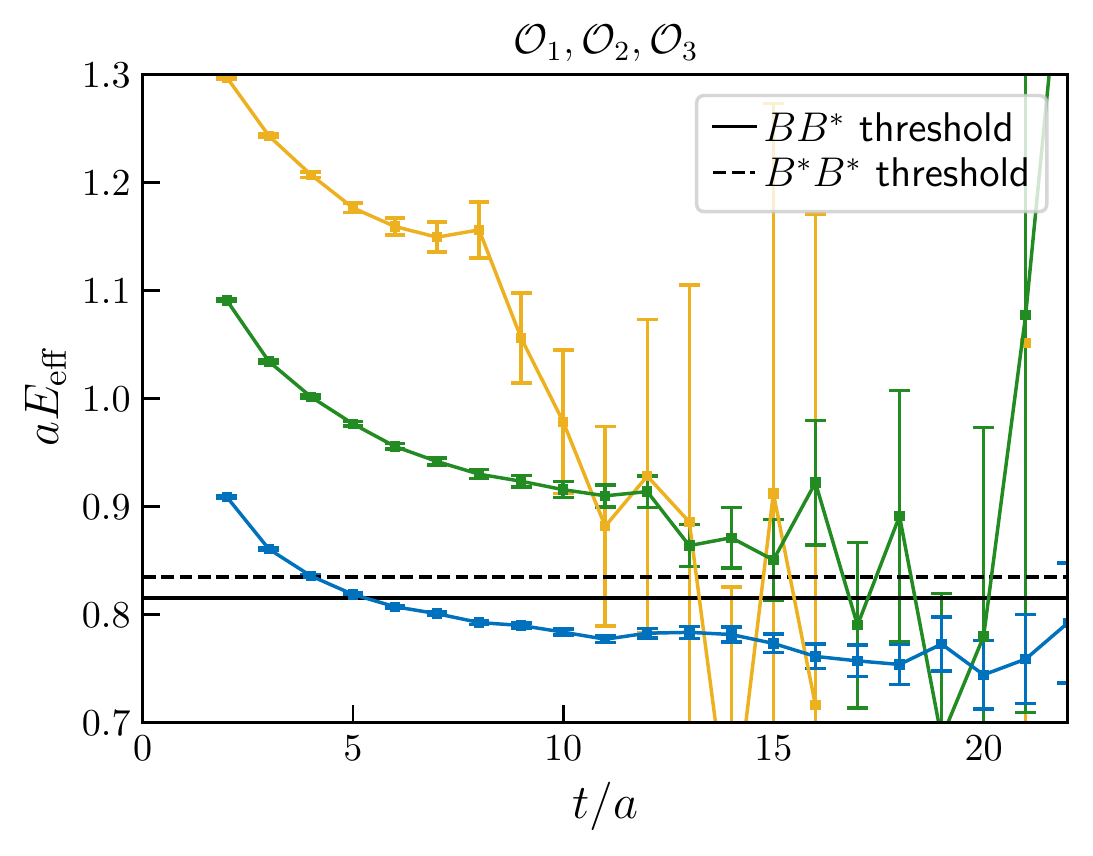}
\includegraphics[width=0.48\textwidth]{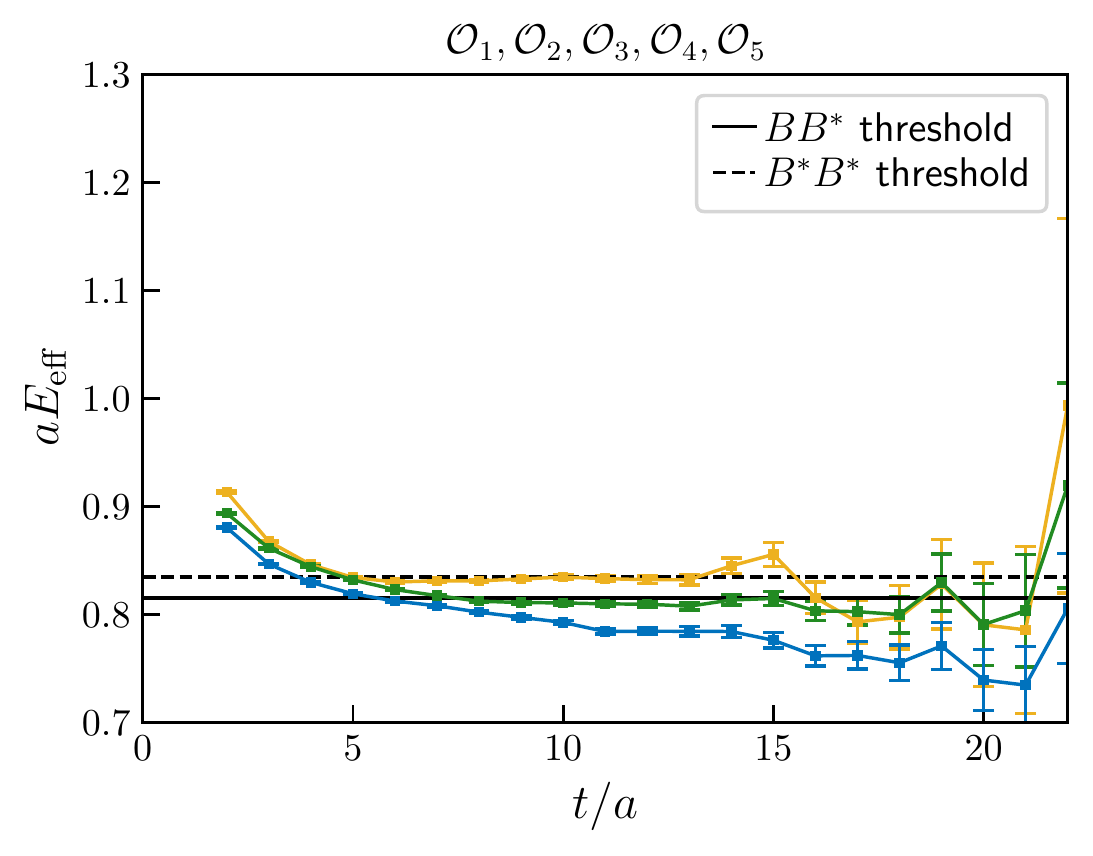}
\end{center}
\caption{\label{FIG001}Effective energies from a GEVP for ensemble a09m310. \textbf{(left)}~$3 \times 3$ matrix, only local interpolating operators $\op_1$, $\op_2$ and $\op_3$. \textbf{(right)}~$5 \times 5$ matrix, both local interpolating operators $\op_1$, $\op_2$ and $\op_3$ and scattering interpolating operators $\op_4$ and $\op_5$.}
\end{figure}

The left plot in Figure~\ref{FIG001} shows effective masses from a GEVP using only local interpolating operators $\op_1$, $\op_2$ and $\op_3$ (i.e.\ corresponding to a $3 \times 3$ matrix). The plateaus exhibit a strong discrepancy with the expectation discussed in the previous paragraph. The right plot in Figure~\ref{FIG001} shows effective masses from a GEVP using both local interpolating operators $\op_1$, $\op_2$ and $\op_3$ as well as scattering interpolating operators $\op_4$ and $\op_5$ (i.e.\ corresponding to a $5 \times 5$ matrix). These effective masses are consistent with the expectation. Thus, Figure~\ref{FIG001} demonstrates that scattering operators are essential for a precise determination of scattering states.

% ********************

\section{Scattering analysis}

To determine the mass of the $\bar b \bar b u d$ tetraquark in infinite volume (for a given ensemble, i.e.\ at given $m_\pi$ and nonzero $a$), we proceed as in our previous work \cite{Leskovec:2019ioa}:
\begin{itemize}
\item[(1)] Compute the two lowest energy levels in the finite spatial volume (see section~\ref{SEC001}).

\item[(2)] Compute the corresponding phase shifts $\delta_0(k_0)$, $\delta_0(k_1)$ using L\"uscher's finite-volume method \cite{Luscher:1990ux}.

\item[(3)] Parameterize $\delta_0(k_0)$, $\delta_0(k_1)$ using the effective-range expansion,
\begin{eqnarray}
\label{EQN001} k \cot(\delta_0(k)) = \frac{1}{a_0} + \frac{r_0}{2} k^2
\end{eqnarray}
with fit parameters $a_0$ and $r_0$.

\item[(4)] The mass of the $\bar b \bar b u d$ tetraquark (and the energy of the first excitation) in infinite spatial volume corresponds to a pole in the scattering amplitude
\begin{eqnarray}
\label{EQN002} T_0(k) = \frac{1}{\cot(\delta_0(k)) - i} .
\end{eqnarray}
The position of the pole can be obtained via Eqs.\ (\ref{EQN001}) and (\ref{EQN002}).
\end{itemize}

\begin{figure}[htb]
\begin{center}
\includegraphics[width=0.43\textwidth,page=1]{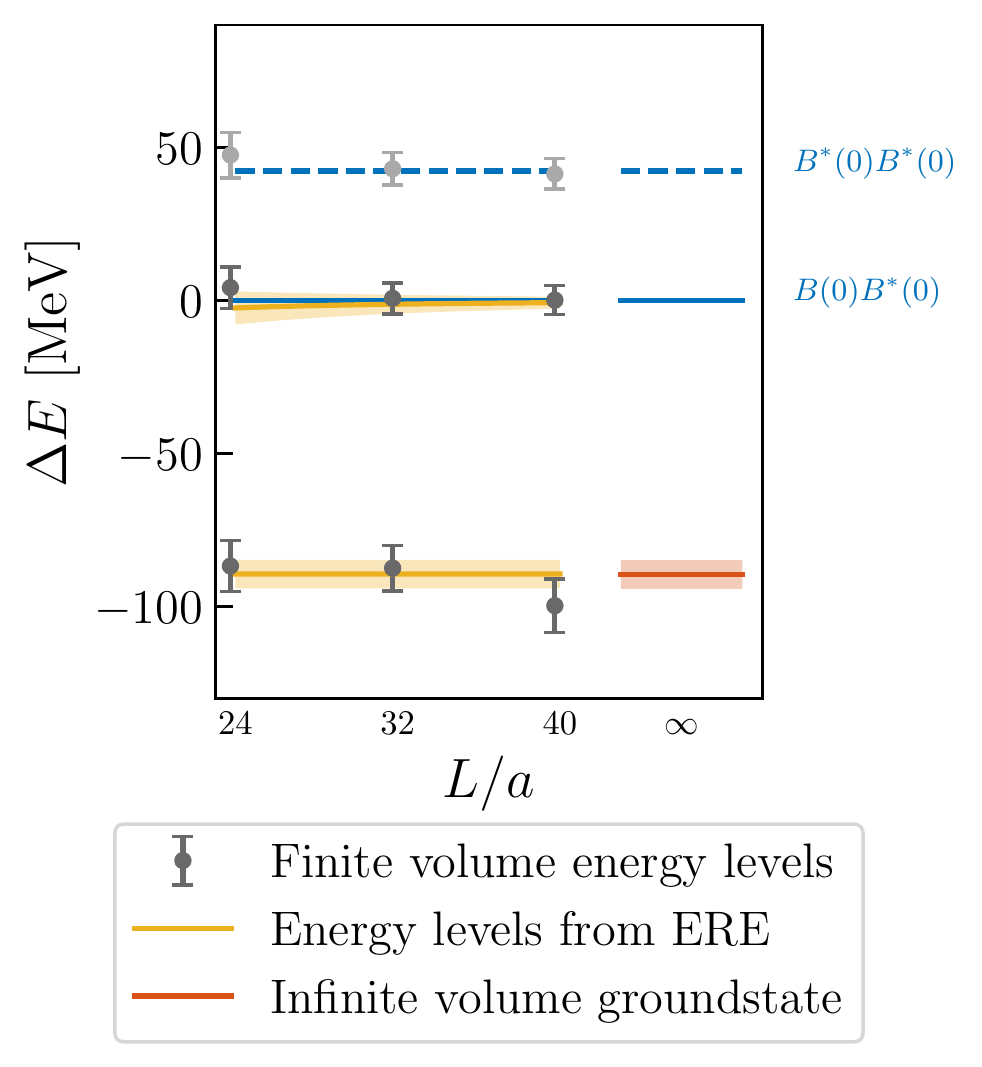}
\includegraphics[width=0.55\textwidth,page=1]{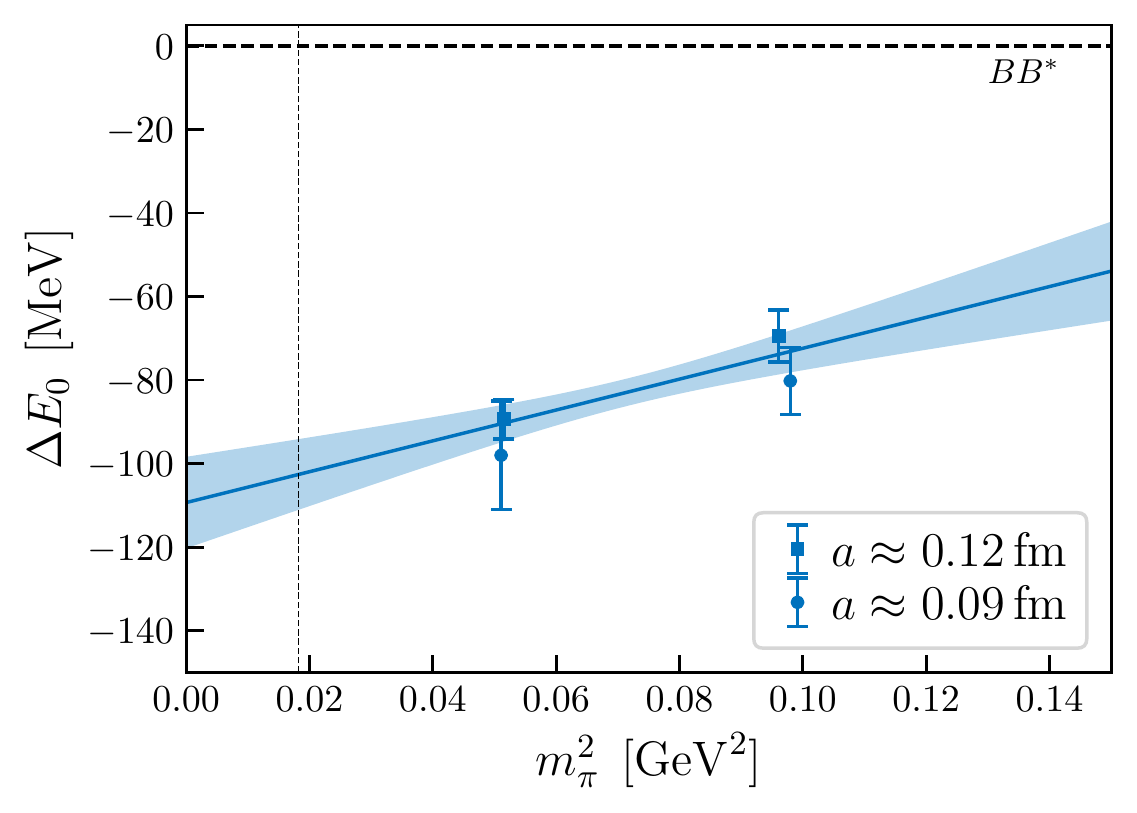}
\end{center}
\caption{\label{FIG002}\textbf{(left)}~Lattice QCD finite volume energy levels (dark gray: ground state and first excitation) for three different volumes $V = L^3$ with $L/a = 24, 32, 40$, but identical $a \approx 0.12 \, \textrm{fm}$ and $m_\pi \approx 220 \, \textrm{MeV}$ (ensembles a12m220S, a12m220, a12m220L) together with a fit based on the effective range expansion~(\ref{EQN001}). \textbf{(right)}~Extrapolation of the tetraquark binding energy in the light $u/d$ quark mass to the physical point.}
\end{figure}

The dark-gray data points in the left plot of Figure~\ref{FIG002} represent lattice QCD finite-volume energy levels (ground state and first excitation) for three different volumes $V = L^3$ with $L/a = 24, 32, 40$, but identical $a \approx 0.12 \, \textrm{fm}$ and $m_\pi \approx 220 \, \textrm{MeV}$ (ensembles a12m220S, a12m220, a12m220L). The orange curves correspond to the two lowest finite volume energy levels as functions of the spatial extent $L$, computed with L\"uscher's finite-volume method using the effective-range expansion~(\ref{EQN001}). There are rather small differences between the finite-volume and infinite-volume energy levels. We attribute this to the large binding energy, $\Delta E_0 \approx \mathcal{O}(100 \, \textrm{MeV})$. Scattering analyses are, however, expected to be more important for smaller binding energies (e.g.\ for the $\bar b \bar b s u$ system with $J^P = 1^+$) and essential for tetraquark resonances (e.g.\ for the $\bar b \bar b u d$ system with $I(J^P) = 0(1^-)$ \cite{JH2022}).

The right plot of Figure~\ref{FIG002} shows an extrapolation of the tetraquark binding energy in the light $u/d$ quark mass based on the six ensembles listed in Table~\ref{TAB001}. The preliminary result at the physical pion mass $m_{\pi,\textrm{\scriptsize phys}} = 135 \, \textrm{MeV}$ is
\begin{eqnarray}
\Delta E_0(m_{\pi,\textrm{\scriptsize phys}}) \approx (-103 \pm 8) \:\textrm{MeV} ,
\end{eqnarray}
where only the statistical uncertainty is shown. This binding energy is slightly smaller than, but consistent with, previous lattice results \cite{Francis:2016hui,Junnarkar:2018twb,Hudspith:2020tdf,Mohanta:2020eed,Leskovec:2019ioa}.

% ********************

\section*{Acknowledgements}

% We acknowledge useful discussions with \textbf{XXXXX ??? XXXXX}

M.W.\ acknowledges support by the Heisenberg Programme of the Deutsche Forschungsgemeinschaft (DFG, German Research Foundation) -- project number 399217702. M.W.\ and M.P.\ acknowledge support by the Deutsche Forschungsgemeinschaft (DFG, German Research Foundation) -- project number 457742095.
J.F.~is financially supported by the H2020 project PRACE 6-IP (GA No.\ 82376) and by the EuroCC project (GA No.\ 951732) funded by the Deputy Ministry of Research, Innovation and Digital Policy and the Cyprus Research and Innovation Foundation and the European High-Performance Computing Joint Undertaking (JU) under grant agreement No 951732.
S.M.\ is supported by the U.S.\ Department of Energy, Office of Science, Office of High Energy Physics under Award Number DE-SC0009913. We thank the MILC collaboration for sharing their gauge-link ensembles \cite{MILC:2012znn}.
Part of the results were obtained obtained using Cyclone High Performance Computer at The Cyprus Institute, under the preparatory access with id \texttt{p054}.
Calculations were conducted on the GOETHE-HLR and on the FUCHS-CSC high-performance computers of the Frankfurt University. We would like to thank HPC-Hessen, funded by the State Ministry of Higher Education, Research and the Arts, for programming advice.
 
% ********************

% ********************

\end{document}